# Time Series Analysis in American Stock Market Recovering in Post COVID-19 Pandemic Period

Weilin Fu , Zhuoran Li , Yupeng Zhang, Xingyou Zhou

Dept. of Statistical Science, University of Toronto./700 University Ave, Toronto, ON M5G 1X6, Canada

*Corresponding author: yp.zhang@mail.utoronto.ca

## ABSTRACT

Every financial crisis has caused a dual shock to the global economy. The shortage of market liquidity, such as default in debt and bonds, has led to the spread of bankruptcies, such as Lehman Brothers in 2008. Using the data for the ETFs of the S&P 500, Nasdaq 100, and Dow Jones Industrial Average collected from Yahoo Finance, this study implemented Deep Learning, Neuro Network, and Time-series to analyze the trend of the American Stock Market in the post-COVID-19 period. LSTM model in Neuro Network to predict the future trend, which suggests the US stock market keeps falling for the post-COVID-19 period. This study reveals a reasonable allocation method of Long Short-Term Memory for which there is strong evidence.

## 1. INTRODUCTION

In the era of the COVID-19 pandemic, the world's economic market has been greatly affected by this natural misfortune. History shows that an economic collapse has accompanied the rise of the financial market in previous years [1]. Various economics studies [2] have analyzed the impacts of the financial crisis on the economy's performance. These studies [3] have indicated that although the potential economy falls back after a financial crisis, potential growth returns or is even higher than its state before the crisis for most economies. However, many external factors, such as quantitative easing and inflation, have interfered with the recovery estimation [4]. Although many scientists and analysts have researched the recovery after an economic crisis, insufficient attention has been paid to the factors causing a crisis. Since many black swan events happened before an economic crisis, it is difficult to predict the time [5]. This paper explores the recovery time of the economy after the financial crisis of 2007–2008 and the financial crisis during the COVID-19 pandemic, especially the performance of the stock market.

The US stock market's recovery has gotten much attention since the financial crisis. Nevertheless, the certainty of the stock market recovery is not guaranteed, considering it is contingent on different variables. According to Luchtenberg & Vu (2015), contagion is transmitted among nations regardless of the country's level of development [6]. This implies that the economic conditions of other countries can influence the shocks in one country's stock market. The authors report that the United States transmits contagion to all countries except China, attributed to the dollar as the international invoicing currency. Besides, the US also received shocks worldwide, not just in particular regions. Therefore, in this research, the stock price will be used as an indicator of the economic condition.

The importance of the international political economy has been shown in 2019 since the global growth rate dropped to a low point during the year. The mounting trade barriers weighed on international business activities, augmenting the cyclical and structural slowdowns already underway. Some countries like the US started quantitive tightening [7], and other countries like Brazil, India, and Turkey are facing the challenge of financial conditions, global inflation, geopolitical tension, and civil unrest, causing a global economic struggle. Both the 2019 and 2008 surges in the global economy confirm the importance of the financial market in the worldwide market that can impact the country's economy, indicating that the influence of external impacts cannot be overlooked in the global economy. Thus, the effect of external factors must be considered to predict the trend during the pandemic and the future movement in the stock market.

# 2. DATA

## 2.1. Data Description

Below are the variables included in the proposed models:

Table 1 Table of Important Variables

| Variables | Description |
|---|---|
| Year | The variable includes years from 2006 to 2022 |
| Stock Price | The opening stock price of each day |

For the stock price, QQQ, S&P, .IXIC, and DJIA were first analyzed, and selected SPX for the final model. The details of each stock index are as below:

Table 2 Table of Analyzed Stock Index

| Stock Index | Description |
|---|---|
| QQQ | An exchange-traded fund that tracks the Nasdaq-100 Index. The Index includes the 100 largest non-financial companies listed on the Nasdaq based on market cap. |
| S&P | A stock index comprised of the 500 largest U.S. publicly traded companies by market capitalization. |
| .IXIC | A stock market index that includes almost all stocks listed on the Nasdaq stock exchange. |
| DJIA | A stock market index that tracks 30 large, publicly owned blue-chip companies trading on the New York Stock Exchange and Nasdaq. |

Table.2 shows the description of the stock index analyzed. S&P is selected for the final model for the following reason: S&P includes stocks from a wide range of industries, which is more representative and comprehensive. QQQ, .IXIC, and DJIA are insufficient and balanced enough to represent the U.S. stock market. QQQ and .IXIC includes a heavier proportion of technology companies, while DJIA emphasizes more on heavy industry. Moreover, unlike S&P, which uses market capitalization, DJIA is price-weighted and does not use weighted arithmetic mean.

## 2.2. Data Visualization

As stated in 3.1. Data Description, four stock indices including S&P, QQQ, .IXIC, and DJIA were first analyzed and visualized, and S&P was selected for the final model. Data wrangling has been conducted before dividing the dataset into training, validating, and testing sets to generate deep learning models. Figure 1, 2, 3, and 4 present a roughly increasing trend in stock price and a similar distribution for all four stock indices, showing that an expansion or recession would similarly influence the stock price. There was a significant decrease from 2008 to 2009 and a gradual increase in the next few years, which is the process of the financial crisis and its recovery. Starting in 2020, there is a similar recession-recovery pattern, which is caused by COVID-19. These two influential financial events will be compared in the following sections, and the previous financial crisis will be used to predict the recovery of economics and the future trend of stock prices.

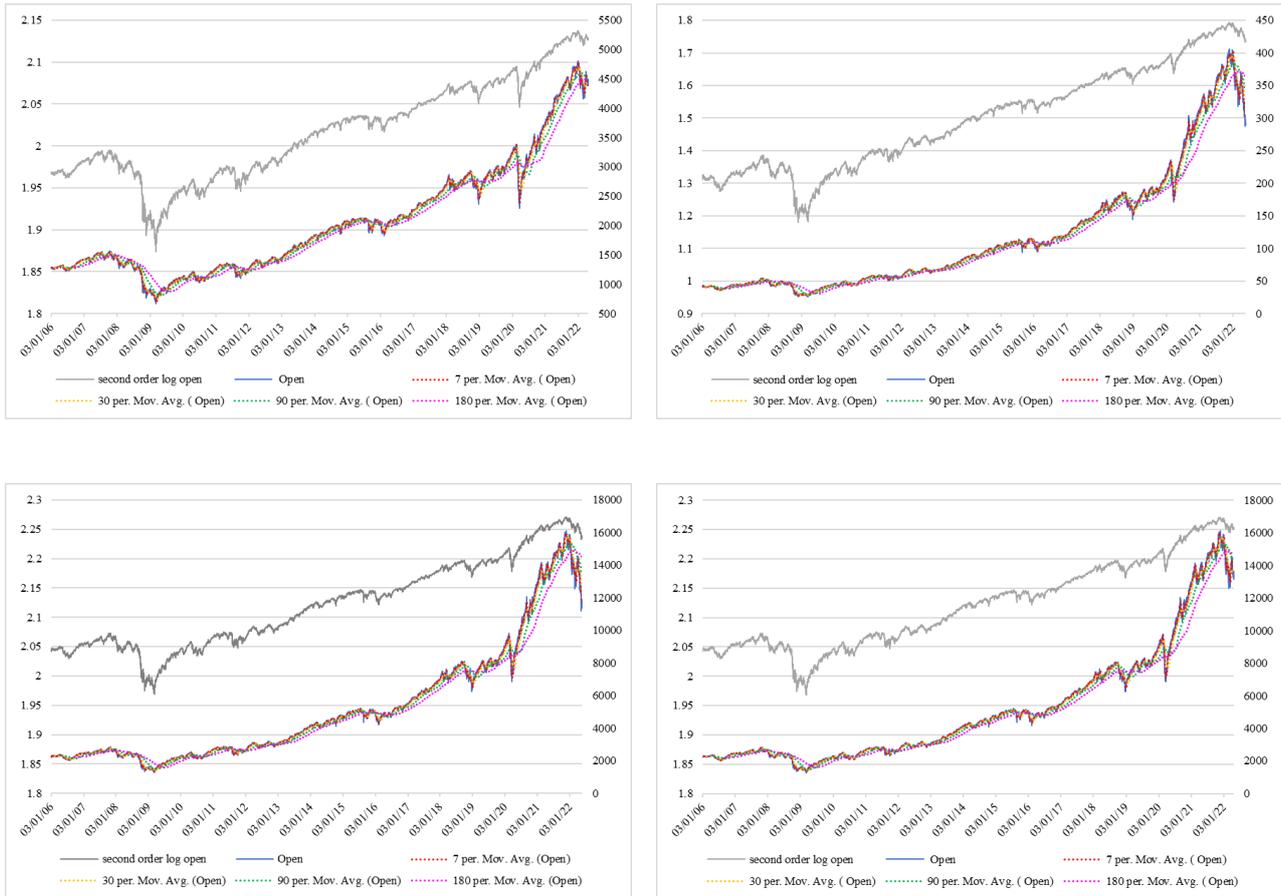

Figure 1 Stock Price and Moving Averages for S&P, QQQ, IXIC , DJIA

## 3. METHOD

### 3.1. ARIMA

The ARIMA model extends AR and MA time series analysis. However, as mentioned in the Data section, there is a trend in this time series. Therefore, this essay will choose the ARIMA model for time series analysis and forecasting. Before applying the model, this time series should satisfy the assumptions: data should be stationary and univariate [8]. Since this project only focuses on stock price, the univariate assumption has been met. To satisfy the stationary assumption, the integrated function in the ARIMA model uses the differencing between observations to make this time series stationary [8]. This study applied the MA model for the MA (moving average) to the differencing moving average. Moreover, the autoregressive model is the model that applies the dependency between observations and lagged observations. The basic ARIMA model is ARIMA (p, d, q), where p indicates the number of lag observations, d shows the degree of differencing, and q is the moving average size. This project chose the function of auto.arima() in R contributed to this result [9]. More code can be seen in the Appendix section.

### 3.2. SARIMA

SARIMA models are extensions of ARIMA models with a seasonal component, and ARIMA fails to encapsulate this information implicitly. By inspecting the dataset, seasonality in the stock price may be found, and applying the SARIMA model would better consider this influential factor [10]. Therefore, this essay applied the stepwise method to find the optimal SARIMA model. The basic SARIMA model is ARIMA (p, d, q)(P, D, Q){m}. The SARIMA method has seven parameters. The first three parameters (p, d, q) are the non-seasonal part, and (P, D, Q){m} deals with the seasonal component of the model, and m is the seasonality period [10]. To define the optimal model, this paper would like to evaluate the model by using values of AIC, BIC, and the model performance matrix. The Model Selection and Performance Table section will further delineate the details.

### 3.3. Model Selection and Performance Table

Akaike Information Criterion (AIC) and Bayesian Information Criterion (BIC) could be applied for evaluating and selecting the ARIMA, SARIMA, and SARIMAX models. AIC is derived from frequentist probability, and BIC is derived from Bayesian probability.

The AIC statistic is defined as below:

$$\text{AIC} = -\frac{2}{N} \times \text{LL} + 2 \times \frac{k}{N} \tag{1}$$

The BIC statistic is defined as below:

$$\text{BIC} = -2 \times \text{LL} + \log(N) \times k \tag{2}$$

In the equations above, N represents the number of objects in the training set, LL is the log-likelihood of the model on the training set, and k is the number of parameters in the model.

Compared to the AIC, the BIC penalizes more complex models, which means it would select simpler models. Therefore, emphasizing BIC more may cause the underfitting problem. Similarly, emphasizing AIC more would cause the problem of overfitting. [11]

This model performance table includes MSE values to evaluate models. For MSE [12], the equation is shown below.

$$\text{MSE} = \frac{1}{n}\sum_{i=1}^{n} (\hat{y}_i - y_i)^2 \tag{3}$$

It presents the distance between the actual and estimated values for each observation. The value of MSE could be 0 to infinity, and the smaller value indicates a more suitable model.

Taking advantage of the collected data and used the Time Series model to find out the patterns. The data that would be used to conduct modeling includes Invesco QQQ Trust Series 1, Standard & Poor's 500, Dow Jones Industrial Average, and NASDAQ Composite. These indexes have a relatively long history and are very stable in the financial industry, even during uncertain periods. Therefore, assuming there are no differences in comparing the 2008 financial crisis and the financial crisis during the COVID-19 pandemic. The recovery time could be patterned and predicted based on the time series model (SARIMA, Seasonal autoregressive integrated moving average model), find some patterns, and predict the recovery time base on the current data set.

To conduct the prediction, divide the current data set into a training, validation, and testing set is necessary. The prospective split ratio would be 7:2:1. Since the data from 2006 to 2010 and the data from the end of 2019 to the current for both indexes has been taken, the data from 2006 to 2010 could be used as the training set and the data from the end of 2019 to the current could be used as validation and testing set.

After the modeling section finished, four independent prediction results could be compared. Whether there exists a significant difference between these results would be explored. The results will be explained in both economic and statistical aspects. Meanwhile, the hypothesis in both aspects according to the results would also be examined and justified.

### 3.4. Long Short-Term Memory (LSTM)

The idea of the Long Short-Term Memory (LSTM) network was proposed by Sepp Hochreiter and Jürgen Schmidhuber in 1997 as an extension of the recurrent neural network (RNN) method. [13] The model of LSTM can be interpreted as follows:

$$\begin{pmatrix} i \\ f \\ o \\ g \end{pmatrix} = \begin{pmatrix} \sigma \\ \sigma \\ \sigma \\ tanh \end{pmatrix} W \begin{pmatrix} h_{t-1} \\ x_t \end{pmatrix} \tag{4}$$

$$c_t = f \odot c_{(t-1)} + i \odot g \tag{5}$$

$$h_t = o \odot tanh(c_t) \tag{6}$$

LSTM is efficient and widely suitable for time series analysis, with various advantages stated below. A common LSTM unit includes a cell and four gates. The forget gate (f) determines whether to erase the cell. The input gate (i) determines whether to write to the cell. The output gate (o) determines how much to reveal the compartment. And the input modulation gate (g) usually determines how much to write to the cell. Backpropagation from $c_t$ to $c_{t-1}$ is an element-wise multiplication. Compared with the full matrix multiplication used in RNN, the speed would greatly increase as the gradient flow would be uninterrupted. Moreover, gradients in LSTM will potentially be multiplying a different forget gate at every time step. Compared with RNN, where the gradient would continuously be multiplied by the same weight matrix, LSTM would greatly relieve the issue of exploding or vanishing gradients in RNN. [14] However, it is gradient-based and the gradient exploding problem still needs to be careful when applying this method. [15] In the study, the S & P 500 data is continuous single cohort data. LSTM is applicable to conduct the time series analysis and prediction on the S & P 500 data. Since LSTM is a time series modeling method with Neural Network, no assumption is required to conduct this modeling. However, the characteristic of LSTM indicates that the data should be a single cohort when training the model. Thus, the open price of the S & P 500 was selected to conduct the modeling. By performing the data wrangling, the discontent in the S & P 500 data has been removed. The S & P 500 data used in the training session is unlabeled and continued, which only contains the open price.

Since LSTM is used as a predictive model in the study, splitting the S & P 500 data into training and testing data would be necessary. The S & P 500 data is generated from 2009.1.1 till 2022.5.20, there are 5984 rows. The method of cross-validation would apply to the LSTM model in the study. To construct the cross-validation, 10% of the total dataset was first split as the test dataset. The split ratio in the study would be 8:2 for the training dataset and validation dataset (Train: Validation: Test = 7:2:1), and the MSE would be considered for comparing the models and conducting further optimizations, the default optimizer is the Root Mean Squared Propagation (RM) algorithm. Using the cross-validation method to compare each model, the drop-out value, the unit, and the lookback would be adjusted to minimize the MSE. Furthermore, the run time is also an important aspect of this study.

The data has been rescaled to (0, 1) before applying to the model. The way to conduct the rescaling process is using the following formula.

$$x_{new} = \frac{x_{original} - x_{min}}{x_{range}} \tag{7}$$

The method used to conduct LSTM would be period-to-point estimation, which uses the S & P 500 data on a rolling basis, like the ARIMA model in the previous section. The lookback period is defined as the previous timesteps the model takes to predict the subsequent timestep. The S & P 500 data is a cohort $X_1, X_2, ..., X_n$, when predicting the value of $X_k$ where $k \in (1, n)$, then the input in the prediction process is based on $X_{k-z}, ..., X_{k-1}$, where z is defined as the lookback value. By using period-to-point estimation, an accurate estimate value would be generated, and the result would be presented in a line plot aside from the true value. Rolling-Window analysis was applied to the LSTM models to discover the trend.

## 4. RESULT

### 4.1. Time Series Analysis

For the ARIMA model, this paper finally decided to use model ARIMA (1, 1, 1) {12} as a result, with AIC value 6578.35, BIC value 6584.77 and MSE value 2.339. For the SARIMA model, Table 3 presents a portion of the stepwise model selection. As mentioned in the Method section, AIC and BIC evaluate performances for each model, and lower AIC and BIC value indicate better model performance. According to the AIC value, SARIMA(1, 2, 1)(0, 1, 1){12} is the most suitable model for forecasting. Differently, according to the BIC values, SARIMA(0, 2, 1)(0, 1, 1){12} is the best model. Then, to ensure the final model has the best performance, this paper decided to use the MSE values of models for the final evaluation. The MSE values are 2.45 and 2.46 models respectively, so by choosing the smaller value, the paper finally decided to take the model SARIMA(1, 2, 1)(0, 1, 1){12}as the final model for forecasting.

Table 3 Process of Optimal SARIMA Model

| Models | AIC | BIC |
|---|---|---|
| ARIMA(0, 2,1)(0, 1, 1){12} | 3103.69 | 3116.52 |
| ARIMA(1, 2,1)(0, 1, 1){12} | 3102.47 | 3119.57 |
| ARIMA(0, 2,1)(1, 1, 1){12} | 3105.63 | 3122.73 |
| ARIMA(2, 2,1)(0, 1, 1){12} | 3104.16 | 3125.65 |
| ARIMA(1, 2,1)(1, 1, 1){12} | 3104.39 | 3125.77 |
| ARIMA(3, 2,1)(0, 1, 1){12} | 3106.70 | 3131.36 |

### 4.2. Long Short-Term Memory (LSTM)

Applying the Long Short-Term Memory (LSTM) technique, various drop-out values, units, and lookback values were tested to optimize the model and minimize the MSE. After setting the drop-out values and the unit, experiments were conducted, and the results are in the following table. The MSE in the following table is based on the training and validation sets. Based on the table below, the final model has been determined and would be used to predict the test set.

Table 4 Comparisons of Look Back Value

| Drop-out | Unit | Lookback | MSE | Runtime |
|---|---|---|---|---|
| 0.2 | 50 | 20 | 0.0097 | 5s (15ms/step) |
| 0.2 | 50 | 50 | 0.00059 | 6s (56ms/step) |
| 0.2 | 50 | 100 | 0.00056 | 9s (233ms/step) |
| 0.2 | 50 | 200 | 0.0026 | 12s (760ms/step) |

Account for the sample size for the test set and the definition of the lookback value. The Long Short-Term Memory (LSTM) model with a lookback value of 50 has a relatively small Mean Squared Error (MSE) and runtime. Additionally, it would keep a reasonable sample size for the test set. Hence, the model with the lookback value of 50 has been determined suitable as the final prediction model based on the runtime and the Mean Squared Error (MSE).

### 4.3. Comparison between Time Series Analysis (SARIMA) and Long Short-Term Memory (LSTM)

Compared the method of Long Short-Term Memory (LSTM) to the Time Series Analysis (SARIMA), Long Short-Term Memory (LSTM) would have a relatively smaller MSE for either the training dataset or the validation dataset. The predicted result and true data based on the test dataset would be plotted in the following to compare the models further. Figure 5 below shows the predicted result and true data based on the test dataset for Long Short-Term Memory (LSTM). The shape of the predicted value is like the true data, except the predicted value is generally more significant than the true data on the training and validation set is generally smaller than the true data on the test data. The MSE is used to measure the prediction error, and the MSE based on the test dataset turns out to be 0.01321, which is reasonable and relatively small compared to the Time Series Analysis (SARIMA) model.

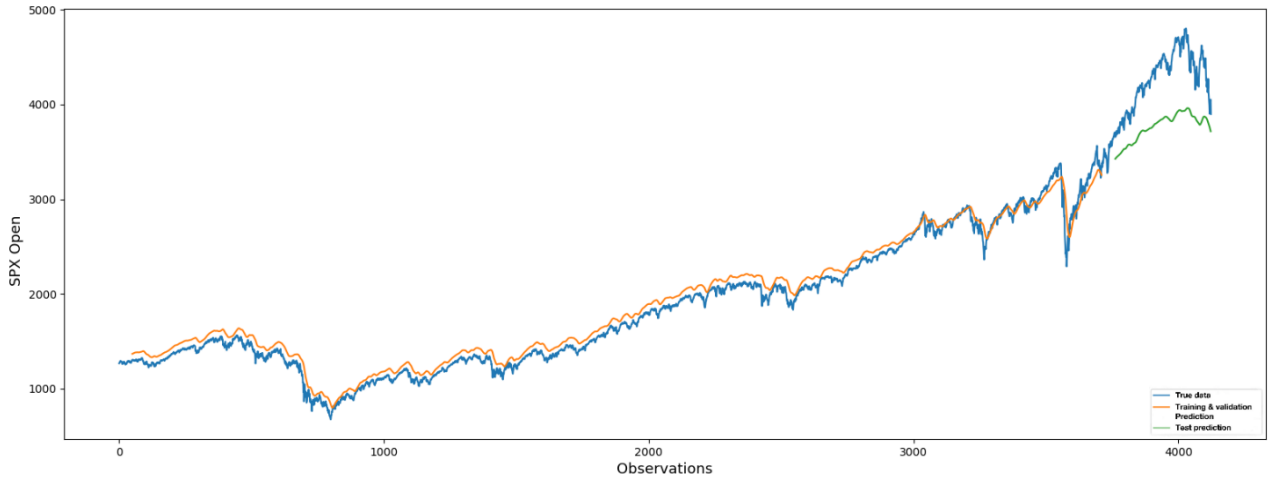

Figure 2 Prediction Curve for LSTM

Figure 6 below presents the prediction result of SARIMA(1, 2, 1)(0, 1, 1){12}. From the plot below, the overall trend of stock price will critically decrease and the range of 95% confidence interval will keep increasing as well. Namely, the result of the SARIMA model indicates that the stock market won't be recovery at least until November 2022, and the accuracy of prediction result is keep reducing.

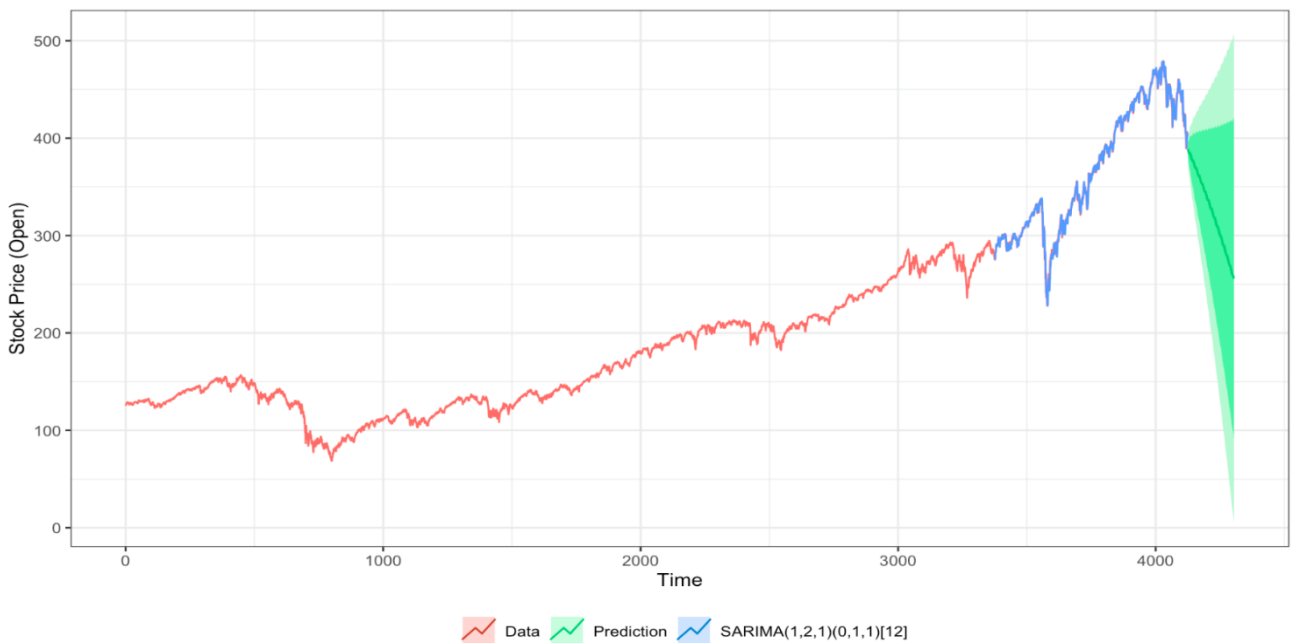

Figure 3 Forecasting Stock Price from 2022.05 to 2022.11

Hence, from the plot and the prediction error above, Long Short-Term Memory (LSTM) would provide a more accurate prediction compared to the Time Series Analysis (SARIMA) model as it has a smaller MSE, and the prediction curve is closer to the true value. Furthermore, the prediction using the Long Short-Term Memory (LSTM) model would be conducted for the next 6 months, from May 2022 to November 2022. The prediction plot is listed below.

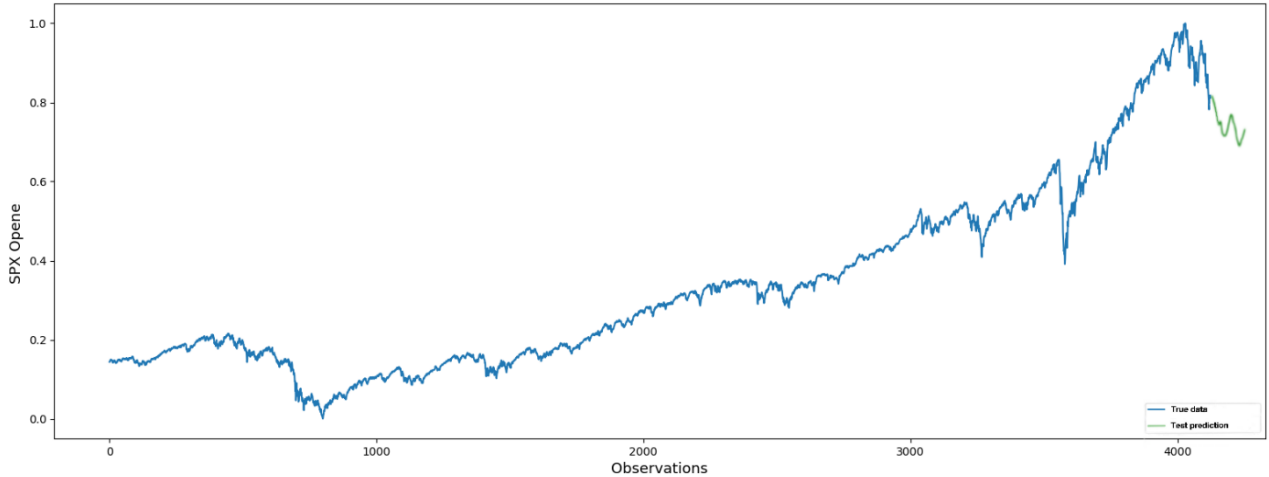

Figure 4 Prediction Curve for LSTM with 6 Months Future Prediction

The plot above shows a concussion in stock price with a downtrend in the forecast of the Long Short-Term Memory (LSTM) model. There would be minor fluctuations in stock prices during this period, but the overall trend would still be declining.

## 5. CONCLUSION

In this research paper, a financial forecasting model for predicting the recovery of the US Stock market during the post-pandemic period has been presented using the SARIMA and LSTM approach. The stock price of S&P, QQQ, DJIA, and IXIC since 2006 has been inspected and the data of S&P 500 was chosen for the model due to it having a more balanced portfolio. In the SARIMA model, the data set is divided into a training and testing set, and the prospective split ratio is 7:3. In the LSTM model, cross-validation was constructed, and the split ratio for the training: validation: testing is approximately also 7:2:1. By analyzing the outcome of the SARIMA and LSTM models, the LSTM model has been identified to have a smaller MSE compared with the model using SARIMA based on the training and validation data set. Hence, LSTM model with lookback value of 50 were chosen, since it has the second smallest MSE, which is 0.00003 larger than the LSTM model with lookback value of 100. However, runtime observed is 6s (56ms/step) when lookback is 50, which is more than four times more efficient than the LSTM model when lookback value of 100. Therefore, LSTM model with lookback value of 50 were chosen to compare with the Time Series model SARIMA(1, 2, 1)(0, 1, 1){12}. Also, this paper could definitionally see that the LSTM model has a better prediction than the Time Series model from the current US stock market trend, which can be found in Appendix. Finally, the model offers a prediction of the economic recovery of the pandemic, and no signals were found that the economy would start to recover in the next 6 months, which is from May 2022 to November 2022. Moreover, the LSTM model in this paper is more predictive as more data becomes available over time.

## 6. DISCUSSION

However, there are some limitations in the prediction model. As stated in Dawei Zhou's research paper, the ARIMA and LSTM approach failed to simultaneously synchronize the alternative data from different sources to financial time series due to the challenges of data heterogeneity. [16] Therefore, the next step of the research would be working on adding data and task heterogeneity to better predict the stock price in different sectors, analyze how pandemics and other economic affairs affect differently, and working on adding data and task heterogeneity to better predict the stock price in different sectors, analyze how pandemics and other economic affairs affect different industries, and optimize the model for future prediction. Furthermore, this essay intends to predict the recovery pattern in the post-covid pandemic period. Hence, further prediction to see the potential recovery pattern could be conducted as the next step.